# Efficient wireless *non-radiative mid-range* energy transfer

Aristeidis Karalis*, J.D. Joannopoulos, and Marin Soljačić

*Center for Materials Science and Engineering
and Research Laboratory of Electronics
Massachusetts Institute of Technology,
77 Massachusetts Avenue, Cambridge, MA 02139, USA*

*Corresponding author:

    Aristeidis Karalis
    Massachusetts Institute of Technology
    77 Massachusetts Avenue, Room 12-104
    Cambridge, MA 02139

    Tel: +1-617-253-6133
    Fax: +1-617-253-2562
    E-mail: aristos@mit.edu

**ABSTRACT**

We investigate whether, and to what extent, the physical phenomenon of long-lifetime resonant electromagnetic states with localized slowly-evanescent field patterns can be used to transfer energy efficiently over non-negligible distances, even in the presence of extraneous environmental objects. Via detailed theoretical and numerical analyses of typical real-world model-situations and realistic material parameters, we establish that such a non-radiative scheme can lead to "strong coupling" between two medium-range distant such states and thus could indeed be practical for efficient medium-range wireless energy transfer.

# I. Introduction

In the early days of electromagnetism, before the electrical-wire grid was deployed, serious interest and effort was devoted (most notably by Nikola Tesla [1]) towards the development of schemes to transport energy over long distances without any carrier medium (e.g. wirelessly). These efforts appear to have met with little success. Radiative modes of omni-directional antennas (which work very well for information transfer) are not suitable for such energy transfer, because a vast majority of energy is wasted into free space. Directed radiation modes, using lasers or highly-directional antennas, can be efficiently used for energy transfer, even for long distances (transfer distance $L_{TRANS} \gg L_{DEV}$, where $L_{DEV}$ is the characteristic size of the device), but require existence of an uninterruptible line-of-sight and a complicated tracking system in the case of mobile objects. Rapid development of autonomous electronics of recent years (e.g. laptops, cell-phones, house-hold robots, that all typically rely on chemical energy storage) justifies revisiting investigation of this issue. Today, we face a different challenge than Tesla: since the existing electrical-wire grid carries energy *almost* everywhere, even a medium-range ($L_{TRANS} \approx few * L_{DEV}$) wireless energy transfer would be quite useful for many applications. There are several currently used schemes, which rely on non-radiative modes (magnetic induction), but they are restricted to very close-range ($L_{TRANS} \ll L_{DEV}$) or very low-power (~mW) energy transfers [2,3,4,5,6].

In contrast to all the above schemes, we investigate the feasibility of using long-lived oscillatory resonant electromagnetic modes, with localized slowly-evanescent field patterns, for efficient wireless *non-radiative mid-range* energy transfer. The proposed method is based on the well known principle of resonant coupling (the fact that two same-frequency resonant objects tend to couple, while interacting weakly with other off-resonant environmental objects) and, in particular, resonant evanescent coupling (where the coupling mechanism is mediated through the overlap of the non-radiative near-fields of the two objects). This well known physics leads trivially to the result that energy can be efficiently coupled between objects in the extremely near field (e.g. in optical waveguide or cavity couplers and in resonant inductive electric transformers). However, it is far from obvious how this same physics performs at mid-range distances and, to our knowledge, there is no work in the literature that demonstrates efficient energy transfer for distances a few times larger that the largest dimension of both objects involved in the transfer. In the present paper, our detailed theoretical and numerical analysis shows that such an efficient mid-range wireless energy-exchange can actually be achieved, while suffering only modest transfer and dissipation of energy into other off-resonant objects, provided the exchange system is carefully designed to operate in a regime of "strong coupling" compared to all intrinsic loss rates. The physics of "strong coupling" is also known but in very different areas, such as those of light-matter interactions [7]. In this favorable operating regime, we quantitatively address the following questions: up to which distances can such a scheme be efficient and how sensitive is it to external perturbations? The omnidirectional but stationary (non-lossy) nature of the near field makes this mechanism suitable for mobile wireless receivers. It could therefore have a variety of possible applications including for example, placing a source (connected to the wired electricity network) on the ceiling of a factory room, while devices (robots, vehicles, computers, or similar) are roaming freely within the room. Other possible applications include electric-engine buses, RFIDs, and perhaps even nano-robots.

## II. Range and rate of coupling

The range and rate of the proposed wireless energy-transfer scheme are the first subjects of examination, without considering yet energy drainage from the system for use into work. An appropriate analytical framework for modeling this resonant energy-exchange is that of the well-known coupled-mode theory (CMT) [8]. In this picture, the field of the system of two resonant objects 1 and 2 is approximated by $F(r,t) \approx a_1(t)F_1(r)+a_2(t)F_2(r)$, where $F_{1,2}(r)$ are the eigenmodes of 1 and 2 alone, and then the field amplitudes $a_1(t)$ and $a_2(t)$ can be shown [8] to satisfy, to lowest order:

$$\frac{da_1}{dt} = -i(\omega_1 - i\Gamma_1)a_1 + i\kappa a_2$$
$$\frac{da_2}{dt} = -i(\omega_2 - i\Gamma_2)a_2 + i\kappa a_1$$
(1)

where $\omega_{1,2}$ are the individual eigenfrequencies, $\Gamma_{1,2}$ are the resonance widths due to the objects' intrinsic (absorption, radiation etc.) losses, and $\kappa$ is the coupling coefficient. Eqs.(1) show that at exact resonance ($\omega_1=\omega_2$ and $\Gamma_1=\Gamma_2$), the normal modes of the combined system are split by $2\kappa$; the energy exchange between the two objects takes place in time $\pi/\kappa$ and is nearly perfect, apart for losses, which are minimal when the coupling rate is much faster than all loss rates ($\kappa \gg \Gamma_{1,2}$) [9]. It is exactly this ratio $\kappa/\sqrt{\Gamma_1\Gamma_2}$ that we will set as our figure-of-merit for any system under consideration for wireless energy-transfer, along with the distance over which this ratio can be achieved. The desired optimal regime $\kappa/\sqrt{\Gamma_1\Gamma_2} \gg 1$ is called "strong-coupling" regime.

Consequently, our energy-transfer application requires resonant modes of high $Q=\omega/2\Gamma$ for low (slow) intrinsic-loss rates $\Gamma$, and this is why we propose a scheme where the coupling is implemented using, not the lossy radiative far-field, but the evanescent (non-lossy) stationary near-field. Furthermore, strong (fast) coupling rate $\kappa$ is required over distances larger than the characteristic sizes of the objects, and therefore, since the extent of the near-field into the air surrounding a finite-sized resonant object is set typically by the wavelength (and quantified rigorously by the "radiation caustic"), this mid-range non-radiative coupling can only be achieved using resonant objects of subwavelength size, and thus significantly longer evanescent field-tails. This is a regime of operation that has *not* been studied extensively, since one usually prefers short tails to minimize interference with nearby devices. As will be seen in examples later on, such subwavelength resonances can often be accompanied with a high radiation-$Q$, so this will typically be the appropriate choice for the possibly-mobile resonant device-object $d$. Note, though, that the resonant source-object $s$ will in practice often be immobile and with less stringent restrictions on its allowed geometry and size, which can be therefore chosen large enough that the near-field extent is not limited by the wavelength (using for example waveguides with guided modes tuned close to the "light line" in air for slow exponential decay therein).

The proposed scheme is very general and *any* type of resonant structure (e.g. electromagnetic, acoustic, nuclear) satisfying the above requirements can be used for its implementation. As examples and for definiteness, we choose to work with two well-known, but quite different, electromagnetic resonant systems: dielectric disks and capacitively-loaded conducting-wire loops. Even without optimization, and despite their simplicity, both will be shown to exhibit acceptably good performance.

### a) Dielectric disks

Consider a 2D dielectric disk object of radius $r$ and relative permittivity $\varepsilon$ surrounded by air that supports high-$Q$ "whispering-gallery" resonant modes (Figure 1). The loss mechanisms for the energy stored inside such a resonant system are radiation into free space and absorption inside the disk material. High-$Q^{rad}$ and long-tailed subwavelength resonances can be achieved, only when the dielectric permittivity $\varepsilon$ is large and the azimuthal field variations are slow (namely of small principal number *m*). Material absorption is related to the material loss tangent: $Q^{abs} \sim \text{Re}\{\varepsilon\}/\text{Im}\{\varepsilon\}$. Mode-solving calculations for this type of disk resonances were performed using two independent methods: numerically, 2D finite-difference frequency-domain (FDFD) simulations (which solve Maxwell's Equations in frequency domain exactly apart for spatial discretization) were conducted with a resolution of *30pts/r*; analytically, standard separation of variables (SV) in polar coordinates was used. The results for two TE-polarized dielectric-disk subwavelength modes of $\lambda/r \geq 10$ are presented in Figure 1. The two methods have excellent agreement and imply that for a properly designed resonant low-loss-dielectric object values of $Q^{rad} \geq 2000$ and $Q^{abs} \sim 10000$ should be achievable.

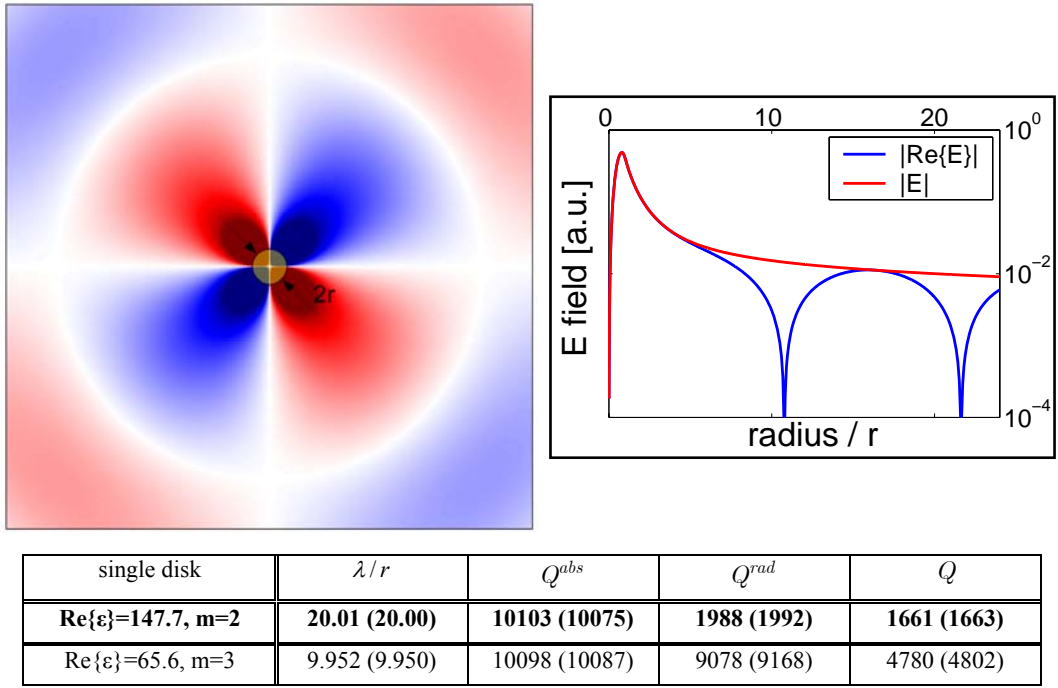

| single disk | $\lambda/r$ | $Q^{abs}$ | $Q^{rad}$ | $Q$ |
|---|---|---|---|---|
| **Re{ε}=147.7, m=2** | **20.01 (20.00)** | **10103 (10075)** | **1988 (1992)** | **1661 (1663)** |
| Re{ε}=65.6, m=3 | 9.952 (9.950) | 10098 (10087) | 9078 (9168) | 4780 (4802) |

**Figure 1:** Main plot: A 2D high-$\varepsilon$ disk of radius *r* (shown in yellow) surrounded by air, along with the electric field (with polarization pointing out of the page) of its resonant whispering-gallery mode superimposed (shown in red/white/blue in regions of positive/zero/negative field respectively). Side plot: Radial plot of the electric field of the mode shown in the main plot (basically a cross-section of the main plot). Note that in air (radius/r>1) the field follows a Hankel-function form, with an initial exponential-like regime (with long tails compared to the small disk size), followed by the oscillatory/radiation regime (whose presence means that energy is slowly leaking out of the disk). Table: Numerical FDFD (and in parentheses analytical SV) results for the wavelength and absorption, radiation and total loss rates, for two different cases of subwavelength-disk resonant modes. Note that disk-material loss-tangent Im{ε}/Re{ε}=10$^{-4}$ was used. *(The specific parameters of the plot are highlighted with **bold** in the table.)* [10]

Note that the required values of $\varepsilon$, shown in Figure 1, might at first seem unrealistically large. However, not only are there in the microwave regime (appropriate for meter-range coupling applications) many materials that have both reasonably high enough dielectric constants and low losses (e.g. Titania: $\varepsilon \approx 96$, $Im\{\varepsilon\}/\varepsilon \approx 10^{-3}$; Barium tetratitanate: $\varepsilon \approx 37$, $Im\{\varepsilon\}/\varepsilon \approx 10^{-4}$; Lithium tantalite: $\varepsilon \approx 40$, $Im\{\varepsilon\}/\varepsilon \approx 10^{-4}$; etc.) [11,12], but also $\varepsilon$ could signify instead the effective index of other known subwavelength ($\lambda/r \gg 1$) surface-wave systems, such as surface-plasmon modes on surfaces of metal-like (negative-$\varepsilon$) materials [13] or metallo-dielectric photonic crystals [14].

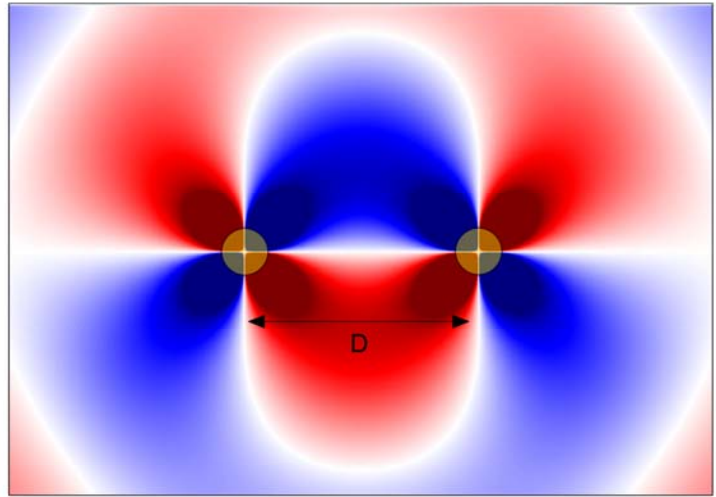

| two disks | $D/r$ | $Q^{rad}$ | $Q=\omega/2\Gamma$ | $\omega/2\kappa$ | $\kappa/\Gamma$ |
|---|---|---|---|---|---|
| **Re{$\varepsilon$}=147.7, m=2** | 3 | 2478 | 1989 | 46.9 (47.5) | 42.4 (35.0) |
| $\lambda/r \approx 20$ | 5 | 2411 | 1946 | 298.0 (298.0) | 6.5 (5.6) |
| $Q^{abs} \approx 10093$ | 7 | 2196 | 1804 | 769.7 (770.2) | 2.3 (2.2) |
|  | **10** | **2017** | **1681** | **1714 (1601)** | **0.98 (1.04)** |
| Re{$\varepsilon$}=65.6, m=3 | 3 | 7972 | 4455 | 144 (140) | 30.9 (34.3) |
| $\lambda/r \approx 10$ | 5 | 9240 | 4824 | 2242 (2083) | 2.2 (2.3) |
| $Q^{abs} \approx 10096$ | 7 | 9187 | 4810 | 7485 (7417) | 0.64 (0.65) |

**Figure 2**: <u>Plot:</u> System of two same 2D high-$\varepsilon$ disks of radius $r$ (yellow) for medium-distance $D$ coupling between them, along with the electric field of the normal mode, which is an even superposition of the single-disk modes of Figure 1, superimposed (red/white/blue). Note that there is also a normal mode, which is an odd superposition of the single-disk modes of Figure 1 (not shown). <u>Table:</u> Numerical FDFD (and in parentheses analytical CMT) results for the *average* of the wavelength and loss rates of the two normal modes (individual values not shown), and also the coupling rate and "strong/weak-coupling" figure-of-merit as a function of the coupling distance $D$, for the two cases of disk modes presented in Figure 1. Only distances for non-radiative ($D < 2r_C$) coupling are considered. Note that the average $\Gamma^{rad}$ (and thus total $\Gamma$) shown are slightly different from the single-disk value of Figure 1, due to far-field interference effects present for the two normal modes, for which CMT cannot make predictions and this is why analytical results for $\Gamma^{rad}$ are not shown but the single-disk value is used. *(The specific parameters of the plot are highlighted with **bold** in the table.)*

To calculate now the achievable rate of energy transfer between two disks 1 and 2, we place them at distance $D$ between their centers (Figure 2). Numerically, the FDFD mode-solver simulations give $\kappa$ through the frequency splitting ($=2\kappa$) of the normal modes of the combined system, which are even and odd superpositions of the initial single-disk modes; analytically, using the expressions for the separation-of-variables eigenfields $\mathbf{E}_{1,2}(\mathbf{r})$ CMT gives $\kappa$ through $\kappa = \omega_1/2 \cdot \int d^3\mathbf{r} \varepsilon_2(\mathbf{r}) \mathbf{E}_2^*(\mathbf{r}) \mathbf{E}_1(\mathbf{r}) \Big/ \int d^3\mathbf{r} \varepsilon(\mathbf{r}) |\mathbf{E}_1(\mathbf{r})|^2$, where $\varepsilon_j(\mathbf{r})$ and $\varepsilon(\mathbf{r})$ are the dielectric functions that describe only the disk j and the whole space respectively. Then, for medium distances $D/r = 10-3$ and for non-radiative coupling such that $D < 2r_C$, where $r_C = m\lambda/2\pi$ is the radius of the radiation caustic, the two methods agree very well, and we finally find (Figure 2) coupling-to-loss ratios in the range $\kappa/\Gamma \sim 1-50$. Although the achieved figure-of-merit values do not fall in the ideal "strong-coupling" operating regime $\kappa/\Gamma \gg 1$, they are still large enough to be useful for applications, as we will see later on.

## b) Capacitively-loaded conducting-wire loops

Consider a loop of radius $r$ of conducting wire with circular cross-section of radius $a$ connected to a pair of conducting parallel plates of area $A$ spaced by distance $d$ via a dielectric of relative permittivity $\varepsilon$ and everything surrounded by air (Figure 3). The wire has inductance $L$, the plates have capacitance $C$ and then the system has a resonant mode, where the nature of the resonance lies in the periodic exchange of energy from the electric field inside the capacitor, due to the voltage across it, to the magnetic field in free space, due to the current in the wire. Losses in this resonant system consist of ohmic loss $R_{abs}$ inside the wire and radiative loss $R_{rad}$ into free space. Mode-solving calculations for this type of *RLC*-circuit resonances were performed using again two independent methods: numerically, 3D finite-element frequency-domain (FEFD) simulations (which solve Maxwell's Equations in frequency domain exactly apart for spatial discretization) were conducted [15], in which the boundaries of the conductor were modeled using a complex impedance $\eta_c = \sqrt{\mu_c \omega / 2\sigma}$ boundary condition, valid as long as $\eta_c/\eta_o \ll 1$ [16] (<*10^-5* for copper in the microwave), where $\mu_o, \varepsilon_o$ and $\eta_o = \sqrt{\mu_o/\varepsilon_o}$ are the magnetic permeability, electric permittivity and impedance of free space and $\sigma$ is the conductivity of the conductor; analytically, the formulas $L = \mu_o r \left[ \ln(8r/a) - 2 \right]$ [17] and $C = \varepsilon_o \varepsilon A / d$, and, in the desired subwavelength-loop ($r \ll \lambda$) limit, the quasi-static formulas $R_{abs} \approx \eta_c \cdot r/a$ (which takes skin-depth effects into account) and $R_{rad} \approx \pi/6 \cdot \eta_o (r/\lambda)^4$ [17] were used to determine the resonant frequency $\omega = 1/\sqrt{LC}$ and its quality factors $Q^{abs} = \omega L / R_{abs}$ and $Q^{rad} = \omega L / R_{rad}$. By tuning the capacitance and thus the resonant frequency, the total $Q$ becomes highest for some optimal frequency determined by the loop parameters: at low frequencies it is dominated by ohmic loss and at high frequencies by radiation. The results for two subwavelength modes of $\lambda / r \geq 70$ (namely highly suitable for near-field coupling and really in the quasi-static limit) at this optimal frequency are presented in Figure 3. The two methods are again in very good agreement and show that expected quality factors in the microwave are $Q^{abs} \geq 1000$ and $Q^{rad} \geq 10000$.

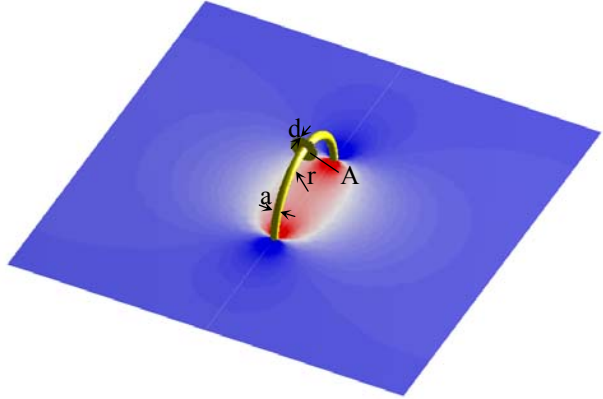

| single loop | $\lambda/r$ | $Q^{rad}$ | $Q^{abs}$ | $Q = \omega/2\Gamma$ |
|---|---|---|---|---|
| **r=30cm, a=2cm ε=10, A=138cm², d=4mm** | **111.4 (112.4)** | **29546 (30512)** | **4886 (5117)** | **4193 (4381)** |
| r=10cm, a=2mm ε=10, A=3.14cm², d=1mm | 69.7 (70.4) | 10702 (10727) | 1545 (1604) | 1350 (1395) |

**Figure 3**: Plot: A wire loop of radius *r* connected to a pair of *d*-spaced parallel plates (shown in yellow) surrounded by air, along with a slice of the magnetic field (component parallel to the axis of the circular loop) of their resonant mode superimposed (shown in red/white/blue in regions of positive/zero/negative field respectively). Table: Numerical FEFD (and in parentheses analytical) results for the wavelength and absorption, radiation and total loss rates, for two different cases of subwavelength-loop resonant modes. Note that for conducting material copper (σ=5.998·10⁷S/m) was used. *(The specific parameters of the plot are highlighted with **bold** in the table.)*

For the rate of energy transfer between two loops 1 and 2 at distance *D* between their centers (Figure 4): numerically, the FEFD mode-solver simulations give $\kappa$ again through the frequency splitting ($= 2\kappa$) of the normal modes of the combined system; analytically, $\kappa$ is given by $\kappa = \omega M / 2\sqrt{L_1 L_2}$, where $M$ is the mutual inductance of the two loops, which, in the quasi-static limit $r \ll D \ll \lambda$ and for the relative orientation shown in Figure 4, is $M \approx \pi/2 \cdot \mu_o (r_1 r_2)^2 / D^3$ [16], which means that $\omega/2\kappa \sim \left(D/\sqrt{r_1 r_2}\right)^3$. Then, and for medium distances $D/r = 10 - 3$, the two methods agree well, and we finally find (Figure 4) coupling-to-loss ratios, which peak at a frequency between those where the single-loop $Q_{1,2}$ peak and are in the range $\kappa/\Gamma \sim 0.5 - 50$.

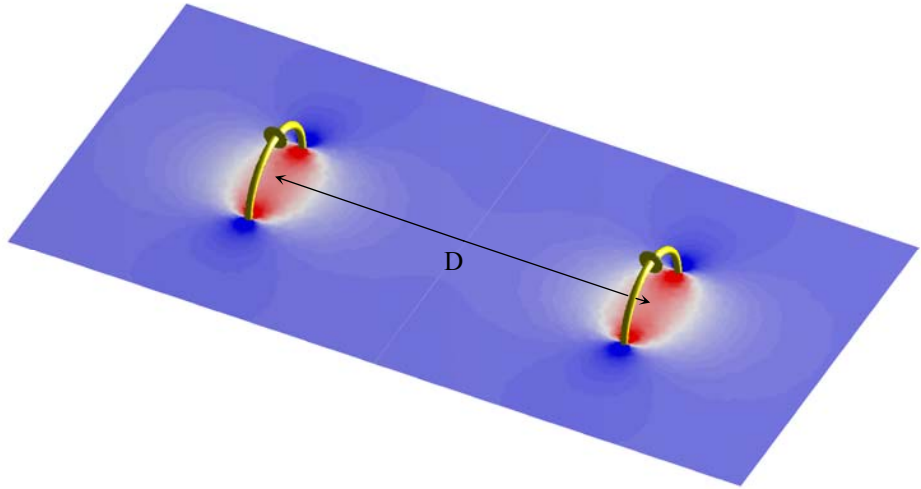

| two loops | $D/r$ | $Q^{rad}$ | $Q=\omega/2\Gamma$ | $\omega/2\kappa$ | $\kappa/\Gamma$ |
|---|---|---|---|---|---|
| **r=30cm, a=2cm** **ε=10, A=138cm², d=4mm** $\lambda/r \approx 112$ $Q^{abs} \approx 4886$ | 3 | 30729 | 4216 | 62.6 (63.7) | 67.4 (68.7) |
| | 5 | 29577 | 4194 | 235 (248) | 17.8 (17.6) |
| | 7 | **29128** | **4185** | **589 (646)** | **7.1 (6.8)** |
| | 10 | 28833 | 4177 | 1539 (1828) | 2.7 (2.4) |
| r=10cm, a=2mm ε=10, A=3.14cm², d=1mm $\lambda/r \approx 70$ $Q^{abs} \approx 1546$ | 3 | 10955 | 1355 | 85.4 (91.3) | 15.9 (15.3) |
| | 5 | 10740 | 1351 | 313 (356) | 4.32 (3.92) |
| | 7 | 10759 | 1351 | 754 (925) | 1.79 (1.51) |
| | 10 | 10756 | 1351 | 1895 (2617) | 0.71 (0.53) |

**Figure 4**: <u>Plot:</u> System of two same wire loops connected to parallel plates (yellow) for medium-distance $D$ coupling between them, along with a slice of the magnetic field of the even normal mode superimposed (red/white/blue). Note that there is also an odd normal mode (not shown). <u>Table:</u> Numerical FEFD (and in parentheses analytical) results for the *average* wavelength and loss rates of the two normal modes (individual values not shown), and also the coupling rate and "strong/weak-coupling" figure-of-merit as a function of the coupling distance $D$, for the two cases of modes presented in Figure 3. Note that the average $\Gamma^{rad}$ shown are again slightly different from the single-loop value of Figure 3, due to far-field interference effects present for the two normal modes, which again the analytical model cannot predict and thus analytical results for $\Gamma^{rad}$ are not shown but the single-loop value is used. *(The specific parameters of the plot are highlighted with **bold** in the table.)*

It is important to appreciate the difference between such a resonant-coupling inductive scheme and the well-known non-resonant inductive scheme for energy transfer. Using CMT it is easy to show that, keeping the geometry and the energy stored at the source fixed, the resonant inductive mechanism allows for ~$Q^2$ (~$10^6$) times more power delivered for work at the device than the traditional non-resonant mechanism. This is why only close-range contact-less medium-power (~W) transfer is possible with the latter [2,3], while with resonance either close-range but large-power (~kW) transfer is allowed [4,5] or, as currently proposed, if one also ensures operation in the strongly-coupled regime, medium-range and medium-power transfer is possible. Capacitively-loaded conductive loops are actually being widely used also as resonant antennas (for example in cell phones), but those operate in the far-field regime with $D/r \gg 1$, $r/\lambda \sim 1$, and the radiation $Q$'s are intentionally designed to be small to make the antenna efficient, so they are not appropriate for energy transfer.

# III. Influence of extraneous objects

Clearly, the success of the proposed resonance-based wireless energy-transfer scheme depends strongly on the robustness of the objects' resonances. Therefore, their sensitivity to the near presence of random non-resonant extraneous objects is another aspect of the proposed scheme that requires analysis. The appropriate analytical model now is that of perturbation theory (PT) [8], which suggests that in the presence of an extraneous object $e$ the field amplitude $a_1(t)$ inside the resonant object 1 satisfies, to first order:

$$\frac{da_1}{dt} = -i(\omega_1 - i\Gamma_1)a_1 + i(\kappa_{11-e} + i\Gamma_{1-e})a_1 \qquad (2)$$

where again $\omega_1$ is the frequency and $\Gamma_1$ the intrinsic (absorption, radiation etc.) loss rate, while $\kappa_{11-e}$ is the frequency shift induced onto 1 due to the presence of $e$ and $\Gamma_{1-e}$ is the extrinsic due to $e$ (absorption inside $e$, scattering from $e$ etc.) loss rate [18]. The frequency shift is a problem that can be "fixed" rather easily by applying to every device a feedback mechanism that corrects its frequency (e.g. through small changes in geometry) and matches it to that of the source. However, the extrinsic loss can be detrimental to the functionality of the energy-transfer scheme, because it cannot be remedied, so the total loss rate $\Gamma_{1[e]} = \Gamma_1 + \Gamma_{1-e}$ and the corresponding figure-of-merit $\kappa_{[e]} / \sqrt{\Gamma_{1[e]}\Gamma_{2[e]}}$, where $\kappa_{[e]}$ the perturbed coupling rate, must be quantified.

## a) Dielectric disks

In the first example of resonant objects that we have considered, namely dielectric disks, small, low-index, low-material-loss or far-away stray objects will induce small scattering and absorption. In such cases of small perturbations these extrinsic loss mechanisms can be quantified using respectively the analytical first-order PT formulas $\Gamma_{1-e}^{rad} \propto \omega_1 \cdot \int d^3\mathbf{r} |\text{Re}\{\varepsilon_e(\mathbf{r})\}|^2 |\mathbf{E}_1(\mathbf{r})|^2 / U$ and $\Gamma_{1-e}^{abs} = \omega_1/4 \cdot \int d^3\mathbf{r} \, \text{Im}\{\varepsilon_e(\mathbf{r})\} |\mathbf{E}_1(\mathbf{r})|^2 / U$, where $U = 1/2 \cdot \int d^3\mathbf{r} \varepsilon(\mathbf{r}) |\mathbf{E}_1(\mathbf{r})|^2$ is the total resonant electromagnetic energy of the unperturbed mode. As one can see, both of these losses depend on the *square* of the resonant electric field tails $\mathbf{E}_1$ at the site of the extraneous object. In contrast, the coupling rate from object 1 to another resonant object 2 is, as stated earlier, $\kappa = \omega_1/4 \cdot \int d^3\mathbf{r} \varepsilon_2(\mathbf{r}) \mathbf{E}_2^*(\mathbf{r}) \mathbf{E}_1(\mathbf{r}) / U$ and depends *linearly* on the field tails $\mathbf{E}_1$ of 1 inside 2. This difference in scaling gives us confidence that, for exponentially small field tails, coupling to other resonant objects should be much faster than all extrinsic loss rates ($\kappa \gg \Gamma_{1-e}$), at least for small perturbations, and thus the energy-transfer scheme is expected to be sturdy for this class of resonant dielectric disks.

However, we also want to examine certain possible situations where extraneous objects cause perturbations too strong to analyze using the above first-order PT approach. For example, we place a dielectric disk $c$ close to another off-resonance object of large $\text{Re}\{\varepsilon\}$, $\text{Im}\{\varepsilon\}$ and of same size but different shape (such as a human being $h$), as shown in Figure 5a, and a roughened surface of large extent but of small $\text{Re}\{\varepsilon\}$, $\text{Im}\{\varepsilon\}$ (such as a wall $w$), as shown in Figure 5b. For distances $D_{h/w}/r = 10-3$ between the disk-center and the "human"-center/"wall", the numerical FDFD simulation results presented in Figure 5 suggest that $Q_{c[h]}^{rad}, Q_{c[w]}^{rad} \geq 1000$ (instead of the initial $Q_c^{rad} \geq 2000$), $Q_c^{abs} \sim 10000$ (naturally unchanged), $Q_{c-h}^{abs} \sim 10^5 - 10^2$, and $Q_{c-w}^{abs} \sim 10^5 - 10^4$, namely the disk resonance seems to be fairly robust, since it is not detrimentally disturbed by the presence of extraneous objects, with the exception of the *very* close proximity of high-loss objects.

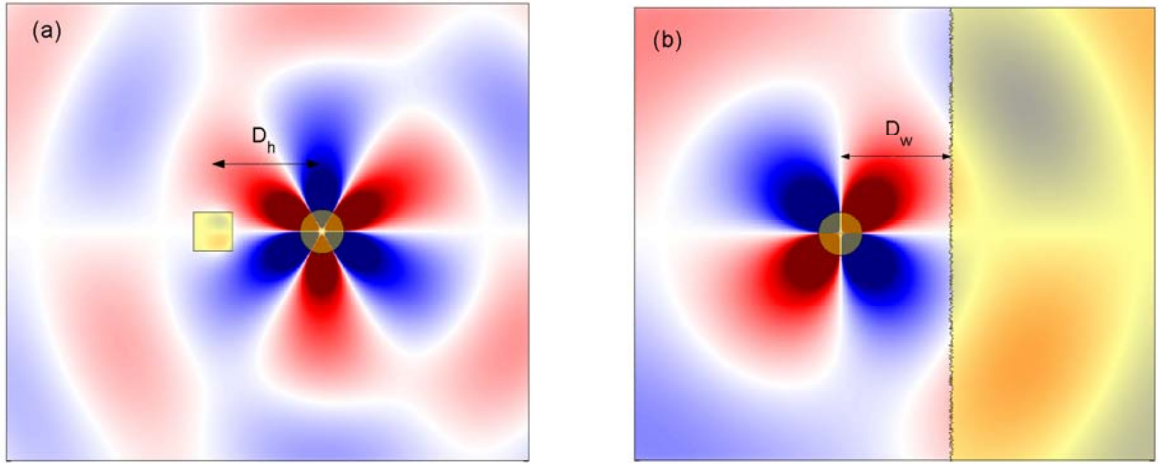

| disk with "human" | $D_h/r$ | $Q_{c-h}^{abs}$ | $Q_{c[h]}^{rad}$ | $Q_{c[h]}$ | disk with "wall" | $D_w/r$ | $Q_{c-w}^{abs}$ | $Q_{c[w]}^{rad}$ | $Q_{c[w]}$ |
|---|---|---|---|---|---|---|---|---|---|
| Re{ε}=147.7, m=2 | 3 | 230 | 981 | 183 | **Re{ε}=147.7, m=2** | 3 | 16725 | 1235 | 1033 |
| $\lambda/r \approx 20$ | 5 | 2917 | 1984 | 1057 | $\lambda/r \approx 20$ | **5** | **31659** | **1922** | **1536** |
| $Q_c^{abs} \approx 10096$ | 7 | 11573 | 2230 | 1578 | $Q_c^{abs} \approx 10098$ | 7 | 49440 | 2389 | 1859 |
|  | 10 | 41496 | 2201 | 1732 |  | 10 | 82839 | 2140 | 1729 |
| **Re{ε}=65.6, m=3** | 3 | 1827 | 6197 | 1238 | Re{ε}=65.6, m=3 | 3 | 53154 | 6228 | 3592 |
| $\lambda/r \approx 10$ | **5** | **58431** | **11808** | **4978** | $\lambda/r \approx 10$ | 5 | 127402 | 10988 | 5053 |
| $Q_c^{abs} \approx 10096$ | 7 | 249748 | 9931 | 4908 | $Q_c^{abs} \approx 10097$ | 7 | 159192 | 10168 | 4910 |
|  | 10 | 867552 | 9078 | 4754 |  | 10 | 191506 | 9510 | 4775 |

**Figure 5**: <u>Plots</u>: A disk (yellow) in the proximity at distance $D_{h/w}$ of an extraneous object (yellow): (a) a high *ε=49+16i* (which is large but actually appropriate for human muscles in the GHz regime [19]) square object of same size (area) with the disk, and (b) a large roughened surface of *ε=2.5+0.05i* (appropriate for ordinary materials such as concrete, glass, plastic, wood [19]), along with the electric field of the disk's perturbed resonant mode superimposed (red/white/blue). <u>Tables</u>: Numerical FDFD results for the parameters of the disk's perturbed resonance, including absorption rate inside the extraneous object and total (including scattering from the extraneous object) radiation-loss rate, for the two cases of disk modes presented in previous Figures. Note that again disk-material loss-tangent Im{ε}/Re{ε}=10[-4] was used, and that $Q_{c[h/w]}^{rad}$ is again different (decreased or even increased) from the single-disk $Q_c^{rad}$ of Figure 1, due to (respectively constructive or destructive) interference effects this time between the radiated and strongly scattered far-fields. (*The specific parameters of the plots are highlighted with* **bold** *in the tables.*)

To examine the influence of large perturbations on an entire energy-transfer system we consider two resonant disks in the close presence of both a "human" and a "wall". The numerical FDFD simulations show that the system performance deteriorates from $\kappa/\Gamma_c \sim 1-50$ (Figure 2) to $\kappa_{[hw]}/\Gamma_{c[hw]} \sim 0.5-10$ (Figure 6), i.e. only by acceptably small amounts.

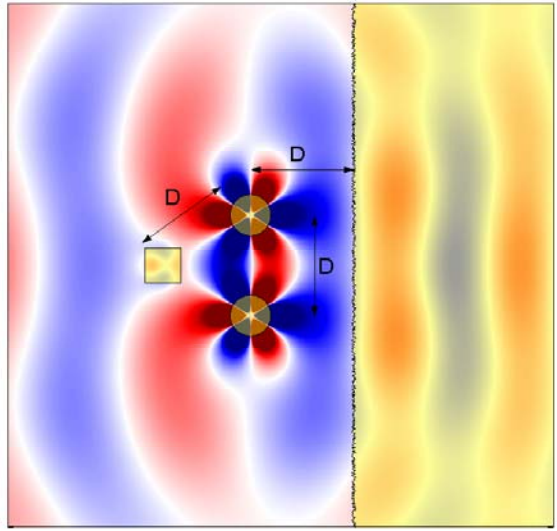

| two disks with "human" and "wall" | $D/r$ | $Q^{abs}_{c-h}$ | $Q^{abs}_{c-w}$ | $Q^{rad}_{c[hw]}$ | $Q_{c[hw]} = \omega/2\Gamma_{c[hw]}$ | $\omega/2\kappa_{[hw]}$ | $\kappa_{[hw]}/\Gamma_{c[hw]}$ |
|---|---|---|---|---|---|---|---|
| Re{ε}=147.7, m=2 | 3 | 3300 | 12774 | 536 | 426 | 48 | 8.8 |
| $\lambda/r \approx 20$ | 5 | 5719 | 26333 | 1600 | 1068 | 322 | 3.3 |
| $Q^{abs}_c \approx 10100$ | 7 | 13248 | 50161 | 3542 | 2097 | 973 | 2.2 |
|  | 10 | 18447 | 68460 | 3624 | 2254 | 1768 | 1.3 |
| **Re{ε}=65.6, m=3** | 3 | 2088 | 36661 | 6764 | 1328 | 141 | 9.4 |
| $\lambda/r \approx 10$ | **5** | **72137** | **90289** | **11945** | **4815** | **2114** | **2.3** |
| $Q^{abs}_c \approx 10100$ | 7 | 237822 | 129094 | 12261 | 5194 | 8307 | 0.6 |

**Figure 6**: <u>Plot:</u> System of two same disks (yellow) for medium-distance $D$ coupling between them in the proximity at equal distance $D$ of two extraneous objects (yellow): both a high $\varepsilon=49+16i$ square object of same size (area) with the disks and a large roughened surface of $\varepsilon=2.5+0.05i$, along with the electric field of the system's perturbed even normal mode superimposed (red/white/blue). <u>Table:</u> Numerical FDFD results for the *average* wavelength and loss rates of the system's perturbed two normal modes (individual values not shown), and also the perturbed coupling rate and "strong/weak-coupling" figure-of-merit as a function of the distance $D$, for the two cases of disk modes presented in previous Figures. Only distances for non-radiative ( $D < 2r_C$ ) coupling are considered. Note once more that the average $\Gamma^{rad}$ takes into account interference effects between all radiated and scattered far-fields. (*The specific parameters of the plot are highlighted with* **bold** *in the table.*)

## b) Capacitively-loaded conducting-wire loops

In the second example of resonant objects that we have considered, the conducting-wire loops, the influence of extraneous objects on the resonances is nearly absent. The reason is that, in the quasi-static regime of operation (*r«λ*) that we are considering, the near field in the air region surrounding the loop is predominantly magnetic (since the electric field is localized inside the capacitor), therefore extraneous non-metallic objects $e$ that could interact with this field and act as a perturbation to the resonance are those having significant magnetic properties (magnetic permeability $Re\{\mu\}>1$ or magnetic loss $Im\{\mu\}>0$). Since almost all every-day materials are non-magnetic, they respond to magnetic fields in the same way as free space, and thus will not disturb the resonance of a conducting-wire loop. To get only a rough estimate of this disturbance, we use the PT formula, stated earlier, $\Gamma_{1-e}^{abs} = \omega_1/4 \cdot \int d^3\mathbf{r}\,\text{Im}\{\varepsilon_e(\mathbf{r})\}|\mathbf{E}_1(\mathbf{r})|^2/U$ with the numerical results for the field of an example like the one shown in the plot of Figure 4 and with a rectangular object of dimensions *30cm x 30cm x 1.5m* and permittivity *ε=49+16i* (human muscles) residing between the loops and almost standing on top of one capacitor (*~3cm* away from it) and find $Q_{c-h}^{abs} \sim 10^5$ and for *~10cm* away $Q_{c-h}^{abs} \sim 5 \cdot 10^5$. Thus, for ordinary distances (*~1m*) and placements (not immediately on top of the capacitor) or for most ordinary extraneous objects *e* of much smaller loss-tangent, we conclude that it is indeed fair to say that $Q_{c-e}^{abs} \to \infty$ and that $\kappa_{[e]}/\Gamma_{[e]} \sim \kappa/\Gamma \sim 0.5 - 50$. The only perturbation that is expected to affect these resonances is a close proximity of large metallic structures.

An extremely important implication of this fact relates to safety considerations for human beings. Humans are also non-magnetic and can sustain strong magnetic fields without undergoing any risk. A typical example, where magnetic fields *B~1T* are safely used on humans, is the Magnetic Resonance Imaging (MRI) technique for medical testing. In contrast, the magnetic near-field required by our scheme in order to provide a few Watts of power to devices is only *B~10⁻⁴T*, which is actually comparable to the magnitude of the Earth's magnetic field. Since, as explained above, a strong electric near-field is also not present and the radiation produced from this non-radiative scheme is minimal, it is reasonable to expect that our proposed energy-transfer method should be safe for living organisms.

In comparison of the two classes of resonant systems under examination, the strong immunity to extraneous objects and the absence of risks for humans probably makes the conducting-wire loops the preferred choice for many real-world applications; on the other hand, systems of disks (or spheres) of high (effective) refractive index have the advantage that they are also applicable to much smaller length-scales (for example in the optical regime dielectrics prevail, since conductive materials are highly lossy).

# IV. Efficiency of energy-transfer scheme

Consider again the combined system of a resonant source $s$ and device $d$ in the presence of a set of extraneous objects $e$, and let us now study the efficiency of this resonance-based energy-transfer scheme, when energy is being drained from the device at rate $\Gamma_{work}$ for use into operational work. The coupled-mode-theory equation for the device field-amplitude is

$$\frac{da_d}{dt} = -i\left(\omega - i\Gamma_{d[e]}\right)a_d + i\kappa_{[e]}a_s - \Gamma_{work}a_d, \tag{3}$$

where $\Gamma_{d[e]} = \Gamma_{d[e]}^{rad} + \Gamma_{d[e]}^{abs} = \Gamma_{d[e]}^{rad} + \left(\Gamma_d^{abs} + \Gamma_{d-e}^{abs}\right)$ is the net perturbed-device loss rate, and similarly we define $\Gamma_{s[e]}$ for the perturbed-source. Different temporal schemes can be used to extract power from the device (e.g. steady-state continuous-wave drainage, instantaneous drainage at periodic times and so on) and their efficiencies exhibit different dependence on the combined system parameters. Here, we assume steady state, such that the field amplitude inside the source is maintained constant, namely $a_s(t) = A_s e^{-i\omega t}$, so then the field amplitude inside the device is $a_d(t) = A_d e^{-i\omega t}$ with $A_d / A_s = i\kappa_{[e]}/(\Gamma_{d[e]} + \Gamma_{work})$. The various time-averaged powers of interest are then: the useful extracted power is $P_{work} = 2\Gamma_{work}|A_d|^2$, the radiated (including scattered) power is $P_{rad} = 2\Gamma_{s[e]}^{rad}|A_s|^2 + 2\Gamma_{d[e]}^{rad}|A_d|^2$, the power absorbed at the source/device is $P_{s/d} = 2\Gamma_{s/d}^{abs}|A_{s/d}|^2$, and at the extraneous objects $P_e = 2\Gamma_{s-e}^{abs}|A_s|^2 + 2\Gamma_{d-e}^{abs}|A_d|^2$. From energy conservation, the total time-averaged power entering the system is $P_{total} = P_{work} + P_{rad} + P_s + P_d + P_e$. Note that the reactive powers, which are usually present in a system and circulate stored energy around it, cancel at resonance (which can be proven for example in electromagnetism from Poynting's Theorem [16]) and do not influence the power-balance calculations. The working efficiency is then:

$$\eta_{work} \equiv \frac{P_{work}}{P_{total}} = \frac{1}{1 + \dfrac{\Gamma_{d[e]}}{\Gamma_{work}} \cdot \left[1 + \dfrac{1}{fom_{[e]}^2}\left(1 + \dfrac{\Gamma_{work}}{\Gamma_{d[e]}}\right)^2\right]}, \tag{4}$$

where $fom_{[e]} = \kappa_{[e]}/\sqrt{\Gamma_{s[e]}\Gamma_{d[e]}}$ is the distance-dependent figure-of-merit of the perturbed resonant energy-exchange system. Depending on the targeted application, reasonable choices for the work-drainage rate are: $\Gamma_{work}/\Gamma_{d[e]} = 1$ (the common impedance-matching condition) to minimize the required energy stored in the source, $\Gamma_{work}/\Gamma_{d[e]} = \sqrt{1 + fom_{[e]}^2} > 1$ to maximize the efficiency for some particular value of $fom_{[e]}$ or $\Gamma_{work}/\Gamma_{d[e]} \gg 1$ to minimize the required energy stored in the device. For any of these choices, $\eta_{work}$ is a function of the $fom_{[e]}$ parameter only. $\eta_{work}$ is shown for its optimal choice in Figure 7 with a solid black line, and is $\eta_{work} > 17\%$ for $fom_{[e]} > 1$, namely large enough for practical applications. The loss conversion ratios depend also on the other system parameters, and the most disturbing ones (radiation and absorption in stray objects) are plotted in Figure 7 for the two example systems of dielectric disks and conducting loops with values for their parameters within the ranges determined earlier.

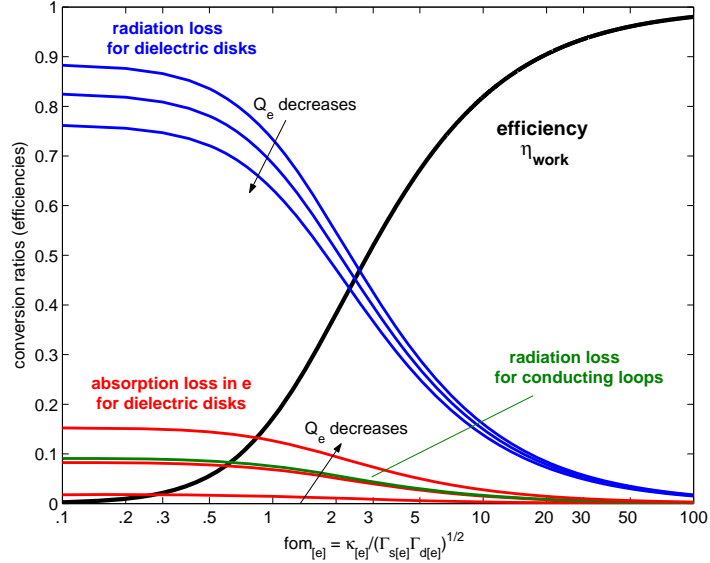

**Figure 7**: <u>Black line:</u> Efficiency of converting the supplied power into useful work ($\eta_{work}$) as a function of the perturbed coupling-to-loss figure-of-merit, optimized with respect to the power-extracting rate $\Gamma_{work}$ (related to the load impedance), for all values of the various quality factors that are present in the system. <u>Blue and red lines:</u> Ratios of power conversion respectively into radiation (including scattering from nearby extraneous objects) and dissipation inside an extraneous object as a function of the figure-of-merit for dielectric disks of $Q_{s[e]}^{rad} = Q_{d[e]}^{rad} \sim 10^3$ and $Q_s^{abs} = Q_d^{abs} \sim 10^4$, and for three values of $Q_{s-e}^{abs} = Q_{d-e}^{abs} = 10^4, 5 \cdot 10^4, 10^5$. <u>Green line:</u> Ratio of power conversion into radiation for conducting-wire loops of $Q_{s[e]}^{rad} = Q_{d[e]}^{rad} \sim 10^4$ and $Q_s^{abs} = Q_d^{abs} \sim 10^3$, and assuming $Q_{s-e}^{abs} = Q_{d-e}^{abs} \to \infty$.

To get a numerical estimate for a system performance, take, for example, coupling distance $D/r = 5$, a "human" extraneous object at distance $D_h/r = 10$, and that $P_{work} = 10W$ must be delivered to the load. Then, for dielectric disks we have (based on Figure 5) $Q_{s[h]}^{rad} = Q_{d[h]}^{rad} \sim 10^3$, $Q_s^{abs} = Q_d^{abs} \sim 10^4$, $Q_{s-h}^{abs} = Q_{d-h}^{abs} \sim 5 \cdot 10^4$ and (based on Figures 2 and 6) $fom_{[h]} \sim 3$, so from Figure 7 we find efficiency $\eta_{work} = 52\%$ and that $P_{rad} \approx 8.3W$ will be radiated to free space, $P_s \approx 0.5W$ will be dissipated inside the source, $P_d \approx 0.3W$ inside the device, and $P_h \approx 0.2W$ inside the human. On the other hand, for conducting loops we have (based on Figures 3 and 4) $Q_{s[h]}^{rad} = Q_{d[h]}^{rad} \sim 10^4$, $Q_s^{abs} = Q_d^{abs} \sim 10^3$, $Q_{s-h}^{abs} = Q_{d-h}^{abs} \to \infty$ and $fom_{[h]} \sim 4$, so we find $\eta_{work} = 61\%$, $P_{rad} \approx 0.6W$, $P_s \approx 3.6W$, $P_d \approx 2.2W$, and most importantly $P_h \to 0$.

## V. Conclusion

In conclusion, we present a scheme based on "strongly-coupled" resonances for mid-range wireless non-radiative energy transfer. Although our consideration has been for a static geometry (namely $\kappa$ and $\Gamma_e$ were independent of time), all the results can be applied directly for the dynamic geometries of mobile objects, since the energy-transfer time $\kappa^{-1}$ ($\sim 1-100\mu s$ for microwave applications) is much shorter than any timescale associated with motions of macroscopic objects. Analyses of very simple implementation geometries provide encouraging performance characteristics and further improvement is expected with serious design optimization. Thus the proposed mechanism is promising for many modern applications. For example, in the macroscopic world, this scheme could potentially be used to deliver power to robots and/or computers in a factory room, or electric buses on a highway (source-cavity would in this case be a "pipe" running above the highway). In the microscopic world, where much smaller wavelengths would be used and smaller powers are needed, one could use it to implement optical inter-connects for CMOS electronics, or to transfer energy to autonomous nano-objects (e.g. MEMS or nano-robots) without worrying much about the relative alignment between the sources and the devices.

As a venue of future scientific research, enhanced performance should be pursued for electromagnetic systems either by exploring different materials, such as plasmonic or metallo-dielectric structures of large effective refractive index, or by fine-tuning the system design, for example by exploiting the earlier mentioned interference effects between the radiation fields of the coupled objects. Furthermore, the range of applicability could be extended to acoustic systems, where the source and device are connected via a common condensed-matter object.

ACKNOWLEDGEMENTS

We should like to thank Prof. John Pendry and L. J. Radziemski for suggesting magnetic and acoustic resonances respectively, and Prof. Steven G. Johnson, Prof. Peter Fisher, André Kurs and Miloš Popović for useful discussions. This work was supported in part by the Materials Research Science and Engineering Center program of the National Science Foundation under Grant No. DMR 02-13282 and by the U.S. Department of Energy under Grant No. DE-FG02-99ER45778.

**REFERENCES**

1) Tesla, N. "Apparatus for transmitting electrical energy." *U.S. patent* number 1,119,732, issued in December 1914.

2) Fernandez, J. M. and Borras, J. A. "Contactless battery charger with wireless control link." *U.S. patent* number 6,184,651, issued in February 2001.

3) Ka-Lai, L., Hay, J. W. and Beart, P. G. W. "Contact-less power transfer." *U.S. patent* number 7,042,196, issued in May 2006. (SplashPower Ltd., www.splashpower.com)

4) Esser, A. and Skudelny, H.-C. "A new approach to power supplies for robots." *IEEE Trans. on industry applications* **27**, 872 (1991).

5) Hirai, J., Kim, T.-W. and Kawamura, A. "Wireless transmission of power and information and information for cableless linear motor drive." *IEEE Trans. on power electronics* **15**, 21 (2000).

6) Scheible, G., Smailus, B., Klaus, M., Garrels, K. and Heinemann, L. "System for wirelessly supplying a large number of actuators of a machine with electrical power." *U.S. patent* number 6,597,076, issued in July 2003. (ABB, www.abb.com)

7) Takao, A. *et al*. "Observation of strong coupling between one atom and a monolithic microresonator.", Nature **443**, 671 (2006).

8) Haus, H. A. *Waves and Fields in Optoelectronics* (Prentice-Hall, New Jersey, 1984).

9) The CMT model is valid exactly for this optimal operational regime of well-defined resonances. Its range of applicability does not include very-close-distance coupling, since there the necessary condition $\kappa \ll \omega_{1,2}$ does not hold, neither large-distance far-field coupling, since it fails to predict far-field interference effects and accurate radiation patterns; rather CMT is exactly suitable for the medium-distance near-field coupling of our interest. Thus the use of this model is justified and the parameters $\kappa, \Gamma_{1,2}$ are well defined.

10) Note that for the 3D case the computational complexity would be immensely increased, while the physics would not be significantly different. For example, a spherical object of $\varepsilon=147.7$ has a whispering gallery mode with $m=2$, $Q_{rad}=13962$, and $\lambda/r=17$.

11) Pozar, D. M. *Microwave Engineering* (Wiley, New Jersey, 2005).

12) Jacob, M. V. "Lithium Tantalate - A High Permittivity Dielectric Material for Microwave Communication Systems." *Proc. of IEEE TENCON 2003* **4**, 1362 (2003).

13) Raether, H. *Surface Plasmons* (Springer-Verlag, Berlin, 1988).

14) Sievenpiper, D. F. *et al.* "3D Metallo-Dielectric Photonic Crystals with Strong Capacitive Coupling between Metallic Islands." *Phys. Rev. Lett.* **80**, 2829 (1998).

15) COMSOL Inc. (www.comsol.com)

16) Jackson, J. D. *Classical Electrodynamics* (Wiley, New York, 1999).

17) Balanis, C. A. *Antenna Theory: Analysis and Design* (Wiley, New Jersey, 2005).

18) The first-order PT model is valid only for small perturbations. Nevertheless, the parameters $\kappa_{11-e}$, $\Gamma_{1-e}$ are well defined, even outside that regime, if $a_1$ is taken to be the amplitude of the exact perturbed mode.

19) Fenske, K. and Misra, D. K. "Dielectric materials at microwave frequencies." *Applied Microwave and Wireless* **12**, 92 (2000).